\documentstyle[prb,epsfig,aps,multicol]{revtex}
\begin{document}
\title{Finite-size corrections to the free energies of crystalline solids}
\author{J. M. Polson\thanks{%
Present address: Physics Dept. McGill University, Montreal, PQ, Canada H3A
2T8.}, E. Trizac\thanks{%
Present address: Laboratoire de Physique Th\'eorique,(UMR 8627 du CNRS), B\^atiment 211, Universit\'e Paris-Sud, 91405 Orsay Cedex (France)}, S. Pronk and D.
Frenkel}
\address{FOM Institute for Atomic and Molecular Physics\\Kruislaan
407\\1098 SJ Amsterdam, The Netherlands}
\date{\today}
\maketitle

\draft
\begin{abstract}
We analyse the finite-size corrections to the free energy of crystals
with a fixed center of mass. When we explicitly correct for the leading ($\ln
N/N$) corrections, the remaining free energy is found to depend
linearly on $1/N$. Extrapolating to the thermodynamic limit
($N\rightarrow\infty$), we estimate the free energy of a defect free
crystal of particles interacting through an $r^{-12}$-potential. We
also estimate the free energy of  perfect hard-sphere crystal near
coexistence: at  $\rho
\sigma^{3}$=1.0409, the excess free energy of a  defect-free
hard-sphere crystal is $5.91889(4)kT$ 
per particle. This, however, is not the free energy of an equilibrium
hard-sphere crystal.  The presence of an finite concentration of
vacancies results in a reduction of the free energy that is some two
orders of magnitude larger than the present error estimate.
\end{abstract}
\begin{multicols}{2}

The earliest numerical technique to compute the free-energy of crystalline
solids was introduced some 30 years ago by  Hoover and Ree~\cite{HooverRee67,HooverRee68}. At present, the ``single-occupancy-cell'' method
of Ree and Hoover is less widely used than the so-called
``Einstein-crystal'' method proposed by Frenkel and Ladd \cite{Frenkel1}.
The latter method employs thermodynamic integration of the Helmholtz free
energy along a reversible artificial pathway between the system of interest
and an Einstein crystal. The Einstein crystals serves as a reference system,
as its free energy can be computed analytically. Since its introduction,
Einstein-crystal method has been used extensively in studies of phase
equilibria involving crystalline solids. For numerical reasons - to suppress
a weak divergence of the integrand - the Einstein-crystal method
calculations have to be carried out at fixed center of mass. The free energy
of the reference crystal is also calculated under the center-of-mass
constraint, and the final calculated free energy of the unconstrained
crystal is determined by correcting for the effect of imposing the
constraint in the calculations. In the original paper, the fixed
center-of-mass constraint was only applied to the particle coordinates, but
not to the corresponding momenta. This is irrelevant as long as one computes
the free-energy difference between two structures that have either both
constrained or both unconstrained centers of \ mass. However, when computing
the absolute free energy of a crystal, one needs to transform from the
constrained to the unconstrained system. In the original paper, this
transformation was not performed consistently. This resulted in a small but
noticeable effect on the computed absolute free energy of the crystal.
Below, we describe the proper approach to calculate the free energy of
arbitrary molecular crystalline solids. The derivation differs from the
earlier work in two respects: first of all, we explicitly show the effect of
momentum constraints. And secondly, we generalize the expression to an
arbitrary crystal containing atoms or molecules with different masses.

The main point of interest involves the calculation of the partition
function of a crystal with and without a constrained center of mass. The
partition function for an unconstrained, $d$-dimensional crystalline solid
of $N_{mol}$ molecules composed of a total of $N$ atoms is given by 
\begin{equation}
Q=c_{N}\int d^{dN}\vec{r}d^{dN}\vec{p}
\exp [-\beta {\cal H}(\vec{r}_{i},\vec{p}_{i})]  \label{eq1} 
\end{equation}
where $c_{N}$=$(h^{dN_{mol}}N_{1}!N_{2}!...N_{m}!)^{-1}$, where there are $N_{1}$ indistinguishable molecules of type 1, $N_{2}$ molecules of type 2,
etc., where $N_{1}+N_{2}+...+N_{m}=N_{mol}$, and $h$ is Planck's constant.
It should be noted that, in all calculations of phase equilibria between
systems that obey classical statistical mechanics, Planck's constant drops
out of the result. Hence, in what follows, we omit all factors $h$. Using the result of the appendix in an article by
Ryckaert and Ciccotti \cite{Ryckaert}, one can show that the constrained
partition function $Q^{con}$ is given by 
\begin{eqnarray}
Q^{con} & = & c_{N}\int d^{dN}\vec{r}d^{dN}\vec{p}\exp [-\beta {\cal H}(\vec{r}_{i},\vec{p}_{i})]  \nonumber \\
&  \times & \delta (\vec{\sigma}(\vec{r}))\delta ({\bf G}^{-1}\cdot 
\overrightarrow{\dot{\sigma}})  \label{eq2}
\end{eqnarray}
where $\vec{\sigma}(\vec{r})$ and $\overrightarrow{\dot{\sigma}}$ are the
constraints and time derivatives of the constraints, respectively, and 
\begin{equation}
G_{ij}=\sum_{k=1}^{N}\frac{1}{m_{k}}\frac{\partial \sigma _{i}}{\partial 
\vec
{r}_{k}}\cdot \frac{\partial \sigma _{j}}{\partial \vec{r}_{k}}
\label{eq3}
\end{equation}
The same integration limits implicit in Eqn.~(\ref{eq1}) are also used in
Eqn.~(\ref{eq2}). To constrain the center of mass (CM), we take
$\vec{\sigma}(\vec{r})=\sum_{i=1}^{N}\mu _{i}\vec{r}_{i}$, and thus, 
$\vec{\dot{\sigma}}=\sum_{i=1}^{N}(\mu _{i}/m_{i})\vec{p}_{i}$, where $\mu _{i}\equiv
m_{i}/\sum_{i}m_{i}$. Note that in Eqns.~(\ref{eq1}) and (\ref{eq2}) we have
assumed that there are no additional internal molecular constraints, such as
fixed bond lengths or bond angles.

We first consider the case of an Einstein crystal, which has a potential
energy function given by $U_{Ein}=(\alpha /2)\sum_{i=1}^{N}(\vec{r}_{i}-\vec{r}_{i}^{(0)})^{2}$, where $\vec{r}_{i}^{(0)}$ are the equilibrium lattice
positions. Note that the particles in a crystal are associated with specific
lattice points and therefore behave as if they are distinguishable -
thus, $c_{N}=1$ (as we omit the factor $1/h^{d(N-1)}$).  It is easy to show that 
\begin{equation}
Q_{Ein}^{CM}=Z_{Ein}^{CM}P_{Ein}^{CM},  \label{eq4}
\end{equation}
with \ 
\begin{eqnarray}
Z_{Ein}^{CM} &=&\int d^{dN}\vec{r}\prod_{i=1}^{N}\exp [-(\beta \alpha
/2)r_{i}^{2}]\delta (\sum_{i=1}^{N}\mu _{i}\vec{r}_{i})\nonumber\\
&=& \left( \frac{\alpha
\beta }{2\pi \sum_{i}\mu _{i}^{2}}\right) ^{d/2}\left( \frac{2\pi }{\alpha
\beta }\right) ^{Nd/2}  \label{eq5}\nonumber \\
&=&\left( \frac{\alpha \beta }{2\pi \sum_{i}\mu _{i}^{2}}\right)
^{d/2}Z_{Ein}
\end{eqnarray}
and 
\begin{eqnarray}
P_{Ein}^{CM} &=&\int d^{dN}\vec{p}\prod_{i=1}^{N}\exp [-(\beta
/2m_{i})p_{i}^{2}]\delta (\sum_{i=1}^{N}\vec{p}_{i})\nonumber\\
&=&\left( \frac{\beta h^{2}}{2\pi M}\right) ^{d/2}\prod_{i=1}^{N}\left( \frac{2\pi m_{i}}{\beta h^{2}}\right) ^{d/2} \nonumber\\
&=&\left( \frac{\beta h^{2}}{2\pi M}\right) ^{d/2}P_{Ein}\text{ ,} \label{eq6}
\end{eqnarray}
where $M=\sum_{i}m_{i}$, while $Z_{Ein}$ and $P_{Ein}$ are the corresponding
contribution to $Q_{Ein}$,the partition function of the unconstrained
Einstein crystal. Clearly, 
\begin{equation}
Q_{Ein}^{CM}=(\sum_{i}m_{i}/\sum_{i}m_{i}^{2})^{d/2}(\beta ^{2}\alpha
h^{2}/4\pi ^{2})^{d/2}Q_{Ein}  \label{eq7}
\end{equation}
Similarly, one can show that the partition function for an arbitrary
crystalline system subject to the CM constraint is given by 
\begin{equation}
Q^{CM}=Z^{CM}(\beta h^{2}/2\pi M)^{d/2}\prod_{i=1}^{N}(2\pi m_{i}/\beta
h^{2})^{d/2},  \label{eq8}
\end{equation}
with

\begin{equation}
Z^{CM}=\int d^{dN}\vec{r}\exp [-\beta U(\vec{r}_{i})]\delta
(\sum_{i=1}^{N}\mu _{i}\vec{r}_{i})  \label{eq10}
\end{equation}
while the partition function of the unconstrained crystal is given by 
\begin{equation}
Q=Z\prod_{i=1}^{N}(2\pi m_{i}/\beta h^{2})^{d/2},  \label{eq9}
\end{equation}
with 
\begin{equation}
Z=\int d^{dN}\vec{r}\exp [-\beta U(\vec{r}_{i})]  \label{eq11}
\end{equation}
Note that, as far as the kinetic part of the partition function is
concerned, the effect of the fixed center of mass constraint {\em is the
same for an Einstein crystal as for an arbitrary ``realistic'' crystal.} 
{\em \ }Using Eqns.~(\ref{eq8}) and (\ref{eq9}), the Helmholtz free energy
difference between the constrained and unconstrained crystal is given by 
\begin{eqnarray}
\beta (F-F^{CM}) &=&-\ln (Q/Q^{CM})  \nonumber \\
&=&\ln (Z^{CM}/Z)-\frac{d}{2}\ln (2\pi M/\beta h^{2})  \label{eq12}
\end{eqnarray}
We note that 
\begin{eqnarray}
\frac{Z^{CM}}{Z} &=&\frac{\int d^{dN}\vec{r}exp[-\beta U(\vec{r}_{i})]\delta \left( \sum_{i}\mu _{i}\vec{r}_{i}\right) }{\int d^{dN}\vec{r}\exp[-\beta U(\vec{r}_{i})]}  \nonumber \\
&=&\langle \delta (\sum_{i}\mu _{i}\vec{r}_{i})\rangle ={\cal P}(\vec{r}_{CM}=\vec{0})  \label{eq13}
\end{eqnarray}
where $\vec{r}_{CM}\equiv \sum_{i}\mu _{i}\vec{r}_{i}$, and ${\cal P}(\vec{r}_{CM})$ is the probability distribution function of the center of mass, $\vec{r}_{CM}$.

To calculate ${\cal P}(\vec{r}_{CM})$ we exploit the fact that the
equilibrium crystal lattice is invariant to translations over displacements
through linear combinations of integer multiples of the lattice vectors.
This is true if the crystal lattice is subject to periodic boundary
conditions. Consequently, the probability distribution of the center of mass
of the lattice is evenly distributed over a volume equal to that of the
Wigner-Seitz cell of the lattice positioned at the center of the volume over
which we carry out the integration in the partition function. Since the
average center of mass of the crystal is equal to the center of mass of the
lattice, it follows that ${\cal P}(\vec{r}_{CM})=1/V_{ws}=N_{ws}/V$, where $V_{ws}$ is the volume of a Wigner-Seitz cell, and $N_{ws}$ is the number of
such cells in the system. Thus, $Z^{CM}/Z={\cal P}(\vec{r}_{CM}=\vec{0})=N_{ws}/V$ In the case of one molecule per cell, this implies $Z^{CM}/Z=N_{mol}/V$, where $N_{mol}$ is the number of molecules in the
system.

In the Frenkel-Ladd free energy calculation, the free energy difference
between the constrained crystal and the reference systems is given by 
\begin{equation}
\beta F^{CM}=\beta F_{Ein}^{CM}-\beta \int_{0}^{1}d\lambda \langle {\Delta U}\rangle _{\lambda }^{CM}  \label{eq14}
\end{equation}
where the statistical average of $\Delta U\equiv U_{Ein}-U$ is calculated by
simulation for fixed CM as a function of $\lambda $ under an effective
potential given by $\tilde{U}(\lambda )=(1-\lambda )U+\lambda U_{Ein}$. Note
that the center of mass must be calculated in the same manner as described
in the paragraph above. Further, note that this expression is only
rigorously valid for systems interacting with continuous potentials. In the
case of particles with discontinuous potentials, e.g. hard particles, the
internal potential energy cannot be turned off continuously. The calculation
for this case differs slightly, and is discussed in detail in the original
article\cite{Frenkel1} and in Ref.~\cite{Frenkel2}.

Using Eqns.~(\ref{eq7}), (\ref{eq12}) and (\ref{eq14}), we find that the free
energy per molecule of the unconstrained crystal is given by 
\begin{eqnarray}
\frac{\beta F}{N_{mol}} &=&-\left( \frac{dN}{2N_{mol}}\right) \ln (2\pi
/\beta \alpha )-\frac{1}{N_{mol}}\ln \prod_{i=1}^{N}\left[ \frac{2\pi m_{i}}{\beta h^{2}}\right] ^{d/2}  \nonumber \\
&&-\frac{\beta }{N_{mol}}\int_{0}^{1}d\lambda \langle {\Delta U}\rangle
_{\lambda }^{CM}  \nonumber \\
&&-\frac{d}{2N_{mol}}\ln \left( \frac{\alpha \beta }{2\pi \sum_{i}\mu
_{i}^{2}}\right) -\frac{\ln (V/N_{mol})}{N_{mol}}  \label{eq15}
\end{eqnarray}

If we consider the special case of a system of single-atom, identical
particles ($m_{i}=m$ and $N=N_{mol}$), we obtain the following: 
\begin{eqnarray}
\frac{\beta F}{N} &=&-\frac{d}{2}\ln \left[ \frac{4\pi ^{2}m}{\alpha \beta
^{2}h^{2}}\right] -\frac{\beta }{N}\int_{0}^{1}d\lambda \langle {\Delta U}\rangle _{\lambda }^{CM}  \nonumber \\
&&-\frac{d}{2N}\ln (\alpha \beta /2\pi )-\frac{d}{2}\frac{\ln N}{N}+\frac{\ln \rho }{N}\text{ ,}  \label{eq16}
\end{eqnarray}
where $\rho \equiv N/V$. The difference between the present result and the
one obtained in ref.\cite{Frenkel1} is in the fourth term on the right-hand
side: $-d\ln N/2N$. The original article implicitly gave the value +$\ln
N/2N $ for a 3D crystal. While the difference between the two expressions
tends to zero in the limit of large $N$, it is non-negligible for system
sizes typically employed in the numerical calculations. However, the
calculated free energy differences between two solids, such as that between
the FCC and HCP hard-sphere crystals, to which the method was applied both
in the original article\cite{Frenkel1} and, more recently, in Ref.~\onlinecite{Frenkel3}, are unaffected by this correction.

In practice, we usually need not calculate the absolute free energy of a
crystal, but excess free energy, $F_{ex}\equiv F-F_{id}$, where $F_{id}$ is
the ideal gas free energy. Let us therefore compute the finite-size
corrections to the latter quantity: \ Given that $\beta F_{id}/N=-\ln
[V^{N}(2\pi m/\beta h^{2})^{dN/2}/N!]/N$, we find 
\begin{eqnarray}
\frac{\beta F_{ex}}{N} &=&-\frac{d}{2}\ln \left[ \frac{2\pi }{\alpha \beta }\right] -\frac{\beta }{N}\int_{0}^{1}d\lambda \langle {\Delta U}\rangle
_{\lambda }^{CM}  \nonumber \\
&&-\frac{d}{2N}\ln (\alpha \beta /2\pi )+\frac{\ln \rho }{N}\nonumber
\\
&&-\frac{d+1}{2}\frac{\ln N}{N}-\ln \rho +1-\frac{\ln 2\pi }{2N},  \label{eq17}
\end{eqnarray}
where we have used $\ln N!\approx N\ln N-N+(\ln 2\pi N)/2$. 

\begin{center}
\begin{figure}
\epsfig{figure=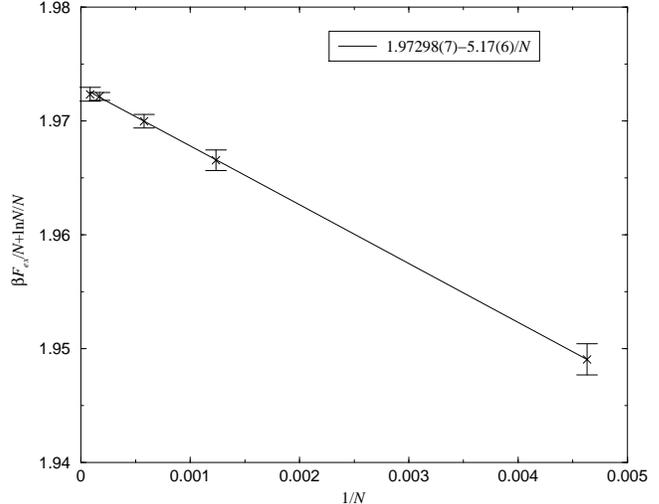,width=6.9cm,angle=270}
\begin{minipage}{7.3cm}
\caption{\label{fig1}
$\beta F_{ex}/N + \ln (N)/N$ vs. $1/N$ for a FCC crystal of soft
($r^{-12}$) spheres at
$k_{B}T/\epsilon $=1.0, and $\rho \sigma ^{3}$=1.1964, 
The solid
line is a linear fit to the data. The coefficient of the $1/N$-term
is: -5.17(6).}
\end{minipage}
\end{figure}
\end{center}

\begin{center}
\begin{figure}
\epsfig{figure=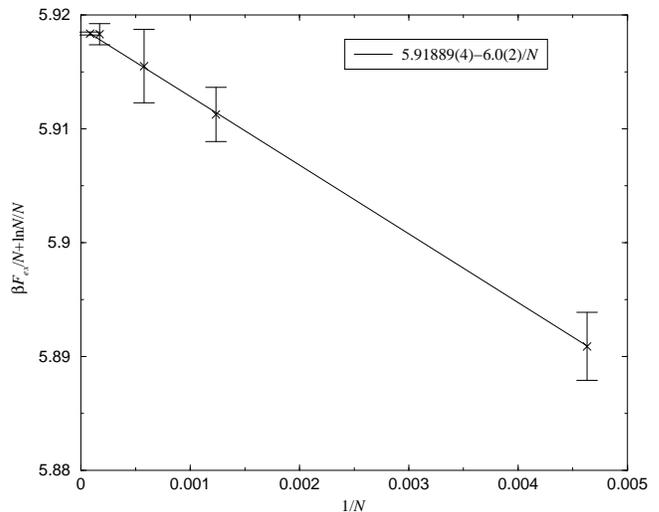,width=6.9cm,angle=270}
\begin{minipage}{7.5cm}
\caption{\label{fig2}
$\beta F_{ex}/N + \ln (N)/N$ vs. $1/N$ for an FCC crystal of hard spheres at
a density $\rho \sigma^{3}$=1.0409.The solid
line is a linear fit to the data. The coefficient of the $1/N$-term
is: -6.0(2).}
\end{minipage}
\end{figure}
\end{center}

Hoover has
analyzed the system-size dependence of the entropy of a classical harmonic
crystal with periodic boundaries\cite{Hoover1}. In this study, it was
established that the leading finite-size correction to the free energy per
particle of a harmonic crystal is equal to $\beta ^{-1}\ln N/N$. If the
harmonic approximation is valid, then this implies that the integral in
Eqn.~(\ref{eq14}) should vary as $+\ln N/N$ plus higher order correction
terms of the order of $N^{-1}$, $N^{-2}$, etc. Consequently, an inspection
of Eqn.~(\ref{eq17}) suggests that $\beta F_{ex}/N+(d-1)\ln N/(2N)$ will
scale as $N^{-1}$, if we neglect terms of order ${\cal O}(1/N^2)$.

To test this prediction we have used the Einstein-crystal method to
calculate the absolute Helmholtz free energy of a (three-dimensional) FCC
crystal of soft spheres interacting with pair potential of $u(r)=\epsilon
(\sigma /r)^{12}$ for systems of size $N$=216, 810, 1728, 5832 and 12096.
The sizes were chosen such that the simulation box shape is cubic for each
system. Further, the simulations were carried out at $k_{B}T/\epsilon $=1.0, 
$\rho \sigma ^{3}$=1.1964, and employed a coupling constant of $\alpha
\sigma ^{2}/\epsilon $=66.0. The results are shown in Figure \ref{fig1} and
are clearly consistent with the predictions. The solid line is a linear fit
which extrapolates to $\beta F_{ex}/N$=1.97298(7)  at $N=\infty $ (where the figure
between brackets is an estimate for the error in the last digit).
Incidentally, we note that, at this density and temperature, the FCC phase of
soft-spheres, is more stable than the HCP phase (by an amount $\Delta
F_{FCC-HCP}/(Nk_BT)=0.0028(8)$). 
The present results suggest that we are  able to correctly account for the
leading ($\ln N/N$) dependence of the free energy of an arbitrary
crystal. In the
analytical calculation of free-energy of a harmonic crystal, it is
always assumed that the center of mass of the crystal 
is fixed. Hence,  the numerical results presented above do not provide
an independent test of the validity of our expression for the contribution
to the free energy due to the center-of-mass motion of the crystal. 

We can perform a similar analysis for a system of hard spheres ($\rho \sigma^{3}$=1.0409, $\lambda_{max}=\alpha/2$=3000). The results are shown in Figure \ref{fig2}.
For hard spheres, $\beta F_{ex}/N$ extrapolates to a value of
$5.91889(4)$  at $N=\infty$, well within the error margin of the original results
of Hoover and Ree\cite{HooverRee68} ($5.924(15)$). Note that the slopes
of the fits (which are proportional to the $1/N$ behavior of the finite
size effect) are similar, although not exactly equal. It should be stressed
that none of these calculations take into account the existence of defects
in the crystal, which, at these levels of precision, is
significant. In fact,  using the early numerical 
results by Bennett and Alder~\cite{BennettAlder} we can estimate 
the equilibrium vacancy concentration in a hard-sphere crystal at
coexistence to be 2.6 $10^{-4}$. Such a vacancy concentration has an
noticeable  effect on the location of the melting point. For instance,
the Gibbs free energy per particle at coexistence is lowered by an
amount $\Delta\mu\approx -3$ $10^{-3}kT$~\cite{Frenkeltobepublished}. 
This correction is far from negligible, as it is some two
orders of magnitude larger than the present numerical accuracy in the absolute
free energy. It is likely that vacancies also lower the
equilibrium free energy of the soft-sphere crystal. However, for that
model, the equilibrium vacancy concentration has, to our knowledge,
not been computed. 

We would like to thank Stella Consta, Benito Groh, Jonathan Doye and Bela
Mulder for stimulating and very useful discussions relating to the material
discussed in this work. This work is part of the research program of the
``Stichting Fundamenteel Onderzoek der Materie'' (FOM) and is supported by
NWO (`Nederlandse Organisatie voor Wetenschappelijk Onderzoek'). JP
acknowledges the financial support provided by the Computational Materials
Science program of NWO, and by the Natural Sciences and Engineering Research
Council of Canada.

\end{multicols}
\end{document}